\documentclass[aps,prx,twocolumn,superscriptaddress,showpacs,floatfix]{revtex4-2}

\usepackage{physics}
\usepackage{graphicx}
\usepackage{dcolumn}
\usepackage{bm}
\usepackage{mhchem}
\usepackage{multirow}
\usepackage{subfiles}
\usepackage{multibib}
\usepackage[colorlinks=true, allcolors=blue]{hyperref}
\usepackage{setspace}

\DeclareUnicodeCharacter{2212}{-}

\begin{document}

\preprint{APS/123-QED}

\title{
Observing the dynamics of octupolar structural transitions in trapped-ion clusters
}

\author{Akhil Ayyadevara}
\affiliation{Raman Research Institute, C. V. Raman Avenue, Sadashivanagar, Bangalore 560080, India}
\author{Anand Prakash}
\affiliation{Raman Research Institute, C. V. Raman Avenue, Sadashivanagar, Bangalore 560080, India}
\author{Shovan Dutta}
\affiliation{Raman Research Institute, C. V. Raman Avenue, Sadashivanagar, Bangalore 560080, India}
\author{Arun Paramekanti}
\affiliation{Department of Physics, University of Toronto, 60 St. George Street, Toronto, ON, M5S 1A7 Canada}
\author{S. A. Rangwala}
\affiliation{Raman Research Institute, C. V. Raman Avenue, Sadashivanagar, Bangalore 560080, India}

\date{\today}

\begin{abstract}

Interacting many-particle systems can self-organize into a rich variety of crystalline structures. While symmetry provides a powerful framework for predicting whether transitions between crystal states are continuous or discontinuous, collective lattice dynamics offer complementary insights into the microscopic mechanisms that drive these transitions. Trapped laser-cooled ions present a pristine and highly controllable few-body system for studying this interplay of symmetry and dynamics. Here, we use real-time fluorescence imaging while deforming the trap potential to observe a variety of structural transitions in three-dimensional (3D), unit-cell-like ion clusters. We identify a set of transitions signaled by parity-odd octupole order parameters, and probe their distinct dynamical signatures. Our observations reveal the softening of a collective Higgs-like mode indicating spontaneous symmetry-breaking, hysteresis resulting from a catastrophe where a metastable state vanishes abruptly, and stochastic switching between metastable states of differing symmetries. We also uncover a remarkable coincidence of symmetry-breaking and discontinuous transitions, analogous to a thermodynamic triple point. Our results establish 3D trapped-ion clusters as a versatile platform to engineer complex potential energy landscapes, opening new avenues for studies of reaction kinetics, geometric frustration, and related phenomena in mesoscopic platforms. 

\end{abstract}

\maketitle

\section{Introduction}

\begin{figure*}[]
    \includegraphics[width=0.97\textwidth]{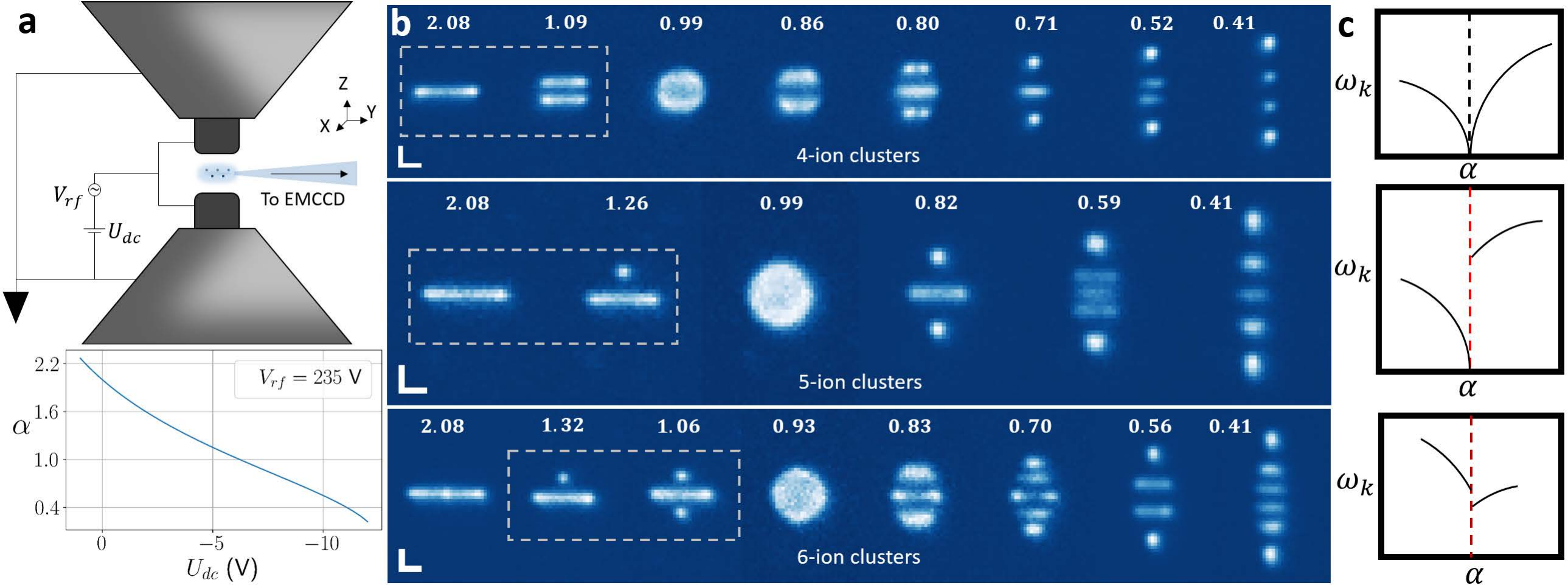}
    \caption{\label{fig:fig1} \textbf{Structural transitions in a 3D \textit{rf} trap}. \textbf{a}, The schematic of the trap and the voltages applied to different electrodes. The outer conical electrodes are grounded, and the inner cylindrical electrodes are driven with a radio-frequency voltage 
    ($V_{rf}$). A DC voltage ($U_{dc}$) is directly applied to the inner electrodes to tune the trap aspect ratio $\alpha\!=\! \omega_z / \omega_x$ as shown in the lower panel (see Methods). The fluorescence is collected along the \textit{y}-axis, and the images are 2D projections onto the \textit{x-z} plane. \textbf{b}, Stitched images of all the observed $4$-, $5$-, and $6$-ion clusters as $\alpha$ is tuned from $\sim2.2$ to $\sim0.25$. The scale bar represents $5$ micrometers. The photon kicks from the Doppler cooling laser apply a stochastic ``torque'' which smears the images along the \textit{x} direction due to free rotation about the cylindrical axis of the trap. \textbf{c}, Characteristic spectrum of the low-lying normal modes across the three transitions of interest identified in the dashed boxes (\textbf{b}). }
\end{figure*}

Phase transitions are ubiquitous in nature, occurring at energy scales ranging from those relevant to solids \cite{CMP_review} and biological systems \cite{Biology_review} to the extreme conditions found in particle colliders and the early universe \cite{Particle_review}. Our understanding of equilibrium phase transitions rests on the foundational framework of Landau theory \cite{Landau_1937}, where symmetry plays a key role in sculpting the free-energy landscape, which dictates the nature of the transition. On the other hand, the real-time transformation from one state to another is governed by dynamical pathways, which can be strikingly different for continuous and first-order transitions. Recent experimental advances in ultrafast diffraction and engineered light-matter interactions have begun to provide insights into transition pathways in dynamics of chemical reactions \cite{ultrafast_chemistry}, atomically resolved crystal dynamics in solid-state materials \citep{ultrafast_THz,ultrafast_weyl,ultrafast_rmp}, and self-organization in atom-cavity systems \cite{Ho2025}. Understanding such transition mechanisms in relation to the underlying symmetry transformations is important for steering molecular reactions \cite{ultrafast_chemistry} and for the dynamical control of quantum materials \cite{ultrafast_weyl, ultrafast_rmp}.

The rich interplay of symmetry and dynamics is exemplified by two different kinds of structural transitions between crystalline states of a solid. Small atomic displacements can trigger a continuous transition to a lower-symmetry configuration, called \textit{displacive} transitions \cite{cochran_book}. These are dynamically signaled by the softening of an optical phonon mode, as observed for the $\alpha$-$\beta$ transition in quartz \cite{Raman1940}. In contrast, crystals can also undergo a discontinuous transition to a state with a different point-group symmetry, accompanied by complex atomic rearrangements \cite{toledano_reconstructive_1996}. The dynamics near such \textit{reconstructive} transitions are characterized by metastability and hysteresis \cite{YAN_hysteresis, AKHANDA_hysteresis} 
due to barriers \cite{toledano_landau_theory} in the free-energy landscape. A single system that can realize a variety of crystalline states with diverse symmetries can provide a unified platform to study and compare the dynamics of distinct crystal transitions.

Trapped-ion Coulomb crystals offer such a versatile and tunable platform \cite{Blatt_sims,drewsen_ion_2015, monroe-revmodphy,sun_two-dimensional_2024, morigi_2025_ioncrystals} 
for experimental investigations of structural phase transitions from few-body clusters \cite{wineland_atomic-ion_1987,dubin_theory_1993, schiffer_phase_1993, mitchell_direct_1998} to mesoscopic ensembles \cite{gilbert_shell-structure_1988, tan_long-range_1995, drewsen_large_1998, mortensen_observation_2006}. The ions can self-organize into a multitude of charge orders depending on the competition between the external confinement and Coulomb repulsion. Fluorescence from Doppler-cooling lasers enables spatially resolved imaging of individual ions and permits the tracking of the real-time crystal configuration. Moreover, the ability to excite and probe collective vibrational modes \cite{ibaraki_parametric_resonance, finite_spectrum_kiethe} enables the investigation of crystal dynamics near structural transitions, which was used to study the linear to zig-zag 
transition in ion chains \cite{finite_spectrum_kiethe, Kiethe2017, zhang_quantum_zigzag_transition}.
In addition, the buckling of 2D Coulomb crystals has been proposed to realize unconventional clock-model transitions
\cite{podolsky_buckling}.

Here, we experimentally probe the dynamics of a wider class of structural transitions in 3D unit-cell-like trapped-ion clusters. By tuning the anisotropy of the confinement potential, we steer the clusters through various regions of stability, uncovering configurations of different symmetries across both displacive and reconstructive transitions. In particular, we identify a set of transitions characterized by parity-odd octupole moments of the cluster; such transformations involving higher-order multipole moments are of current research interest in solid-state physics \cite{Harter2017Science, Ning2023_NatCommQuadrupolar, voleti_octupolar_order} and nuclear physics \cite{Casten2006_nuclear, Rajbanshi2021_nuclear, Zhao2024_nuclearshapeLHC}. We directly observe a Higgs-like mode softening, bistability, hysteresis, as well as a triple-point-like feature, highlighting the complex energy landscapes realizable in 3D ion clusters.

\section{Results}

\subsection{Structural transitions}

We trap clusters of laser-cooled $^{40}\text{Ca}^+$ ions \cite{Anand_oven} in an end-cap type radio-frequency (\textit{rf}) trap \cite{schrama_novel_1993} as shown in Fig.~\ref{fig:fig1} (\textbf{a}). The trap provides a time-averaged harmonic confinement in 3D with cylindrical symmetry (see Methods). The competition between Coulomb interactions and the harmonic confinement determines the equilibrium Configuration(s) of Minimum Energy (CME(s)) \cite{dubin_trapped_1999}, which minimize
\begin{equation}
    V =  \sum_{\substack{i = 1}}^n \Bigl( \frac{m \omega_{x}^2}{2} \, (x_{i}^2+y_{i}^2+\alpha^{ 2}z_{i}^2) + \sum_{\substack{j > i}}^n \frac{k_e e^2}{|\boldsymbol{r}_{i}- \boldsymbol{r}_{j}|}  \Bigr),
\end{equation}
where $\bm{r}_{i} \equiv (x_{i},y_{i},z_{i})$, $i = 1, 2, ...,n$, are the coordinates of the ions with mass \textit{m}, $k_e=1/4\pi \epsilon_0$ is the Coulomb constant, \textit{e} is the elementary charge, $\omega_x$ and $\omega_z$ are the secular frequencies along azimuthal and axial directions, respectively, and $\alpha = \omega_z/\omega_x$ is the aspect ratio of the 3D potential. In the experiment, we tune the trap aspect ratio
$\alpha$ by changing a superimposed DC voltage $U_{dc}$, as shown in Fig.~\ref{fig:fig1} (\textbf{b}). We track the
resulting sequence of configurations and their transitions for $4$-, $5$-, and $6$-ion clusters via fluorescence imaging which captures
2D projections of the cluster configurations.
Due to the cylindrical symmetry of the trap, these ion clusters freely rotate around the $z$-axis over the time of our
imaging, driven by photon kicks from the
Doppler cooling laser.
For $\alpha=1$, the trapping potential becomes completely isotropic, and 
the ion clusters 
undergo free spherical rotation.
As seen from Fig.~\ref{fig:fig1}, for $\alpha\gg1$ the ions assume 2D configurations in the \textit{x-y} plane,
while decreasing $\alpha$ leads to transitions into various 3D structures, until 
the clusters eventually form a linear 1D chain along \textit{z-}axis for $\alpha\ll1$.

\begin{figure*}[]
    \includegraphics[width=0.97\textwidth]{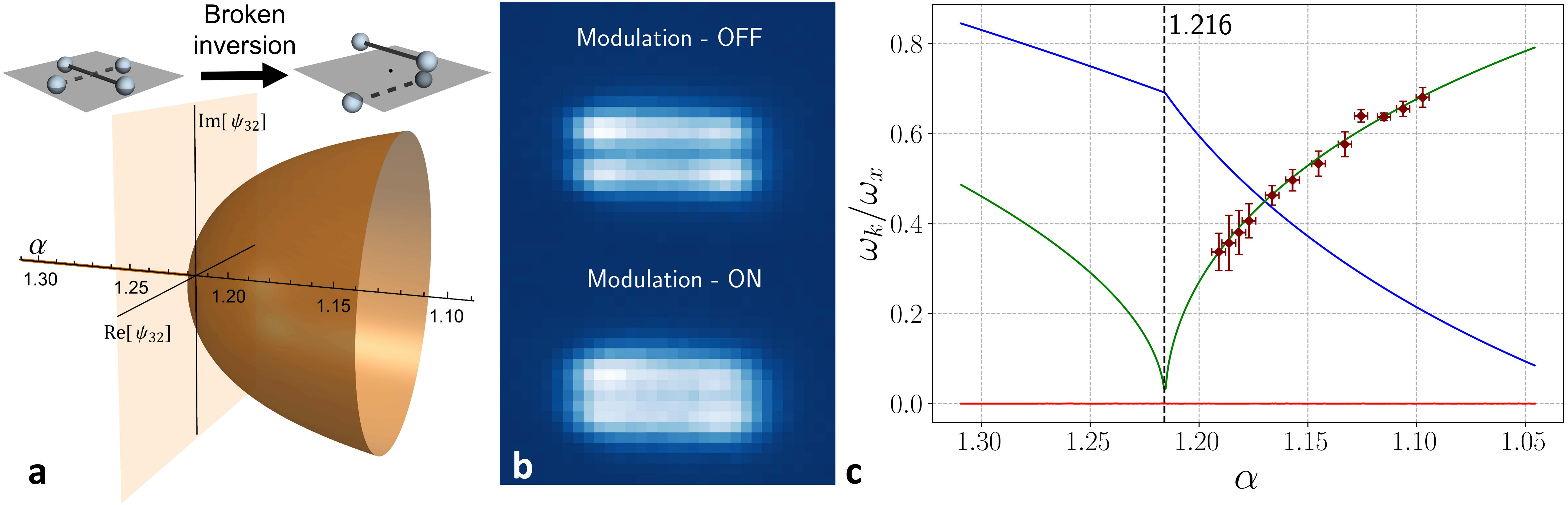}
    \caption{\label{fig:4ion-sm} \textbf{Resonant excitation of a Higgs-like soft mode in $4$-ion tetrahedron.} \textbf{a}, The square to tetrahedron transition visualized as a supercritical pitchfork bifurcation. The order parameter is a complex two-component octupole moment $\psi_{32}$ (see text). \textbf{b}, Response of the tetrahedral cluster to trap voltage modulation captured by fluorescence images. \textbf{c}, The spectra of low-frequency modes computed around the critical point, scaled by the azimuthal secular frequency $\omega_x$. The red line corresponds to the zero-frequency free rotation about the cylindrical axis of the trap. The blue curve corresponds to two degenerate rotational modes about the x- and y-axes. The green curve corresponds to the inversion symmetry-breaking (restoring) mode of the square (tetrahedron), softening at $\alpha\approx1.216$. The experimentally measured Higgs-like excitation on the tetrahedron side falls on top of the computed spectrum. Vertical error bars represent the uncertainty in the peak response frequency of the cluster. Horizontal error bars represent uncertainty in the value of $\alpha$ measured by exciting the motion of a single ion (see Methods).}
\end{figure*}

To characterize the observed transitions, we have numerically obtained the CMEs for $4$-, $5$-, and $6$-ion clusters, and their normal modes at the theoretical limit of zero temperature. We show these results in the supplementary material (Figures SF4, SF5, and SF6). Our numerically obtained CMEs match all the experimentally observed structures and accurately determine their transition points. In the rest of this paper, we focus on the regime of $\alpha \!>\! 1$, unearthing interesting dynamics that emerge near three distinct types of transitions highlighted by dashed boxes in Fig. \ref{fig:fig1} (\textbf{a}). The normal mode features shown in Fig. \ref{fig:fig1} (\textbf{c}) indicate: (i) a symmetry breaking transition in a $4$-ion cluster with mode softening on both sides of the critical point, (ii) a coincidence of buckling instability of a 2D configuration and a symmetry-changing transition to a 3D configuration in a $5$-ion cluster, and (iii) a discontinuous transition in a $6$-ion cluster, where configurations of two different symmetries are bistable over an extended region.

\subsection{Mode softening}

For $\alpha\!\gg\! 1$, four ions self-organize at the corners of a square in the \textit{x-y} plane. As the axial confinement is made weaker by lowering $\alpha$, the square transitions continuously to a tetrahedral arrangement. The two diagonals of the square, represented by the solid and the dashed \textit{pseudo-}bonds in Fig.~\ref{fig:4ion-sm}(\textbf{a}), \textit{displace} along the axial direction. This breaks inversion symmetry about the trap center at the critical point $\alpha_c\!\approx\!1.216$ (see Supplementary discussion I), realizing a supercritical pitchfork bifurcation \cite{strogatz_nonlinear_2015}. 
The lowest-order multipole that captures this inversion-breaking transition is the (complex) parity-odd octupole moment $\psi_{3 2} \!\equiv\! \int d{\bf r} \rho_e({\bf r}) Y_{3,2}({\bf r})$, where $Y_{3,2}({\bf r}) \!\propto\! (x+iy)^2z$ and $\rho_e({\bf r})$ is the charge distribution.

\begin{figure*}[]
    \includegraphics[width=0.97\textwidth]{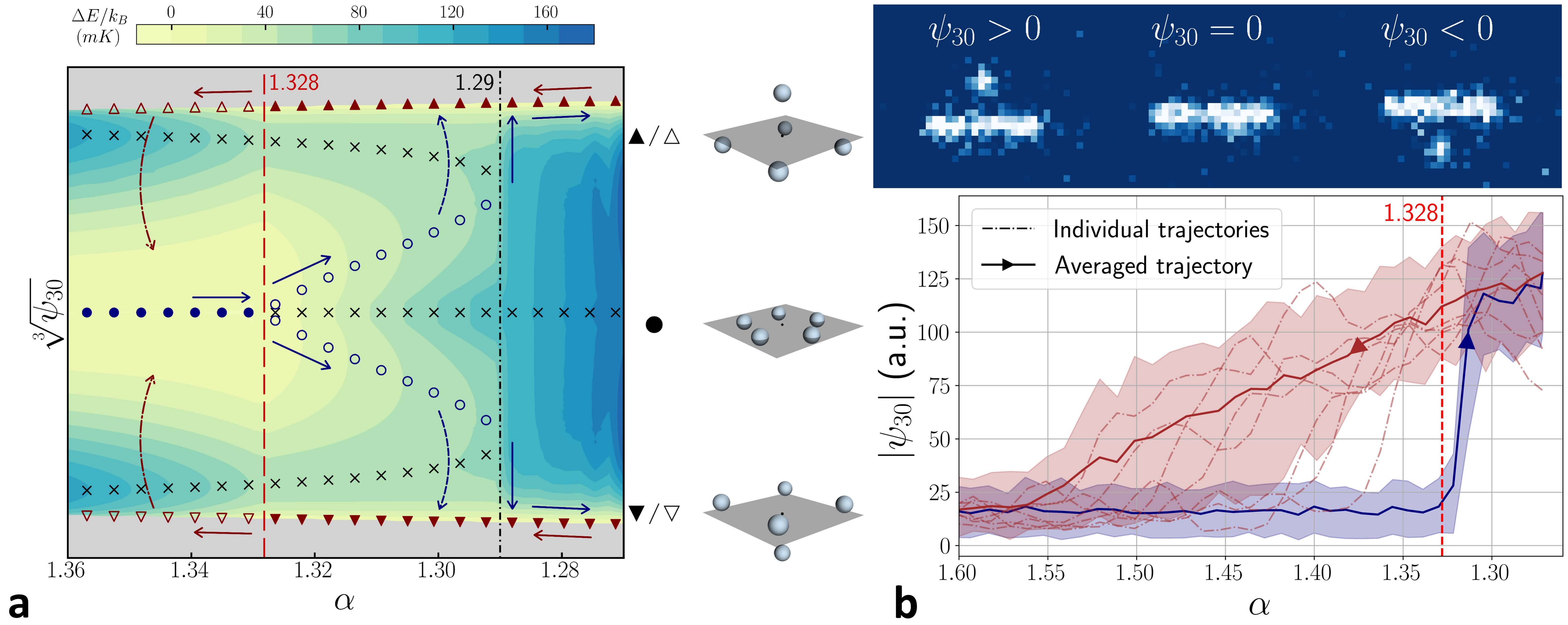}
    \caption{\label{fig:5ion-hys} \textbf{Hysteresis and a triple point in $5$-ion structural transition landscape.} \textbf{a}, The numerically computed energy landscape around the planar to 3D transition in a $5$-ion cluster as a function of $\alpha$ and the octupole order parameter $\psi_{30}$ (see text). We show the minimum energy path for each value of $\alpha$. For a clearer visual representation, we use $\sqrt[3]{\psi_{30}}$ rather than $\psi_{30}$ for the $y$-axis. The solid symbols represent global minima, the open symbols represent metastable minima, and the crosses represent saddles. $\alpha_t \!=\! 1.328$ marks a discontinuous change in the global minimum between the pentagon and the two symmetric square pyramids. Simultaneously, the pentagon undergoes a symmetry-breaking pitchfork bifurcation into two buckled states. These buckled pentagons vanish at $\alpha \!\approx\! 1.29$ via saddle-node bifurcations. The arrows mark the \textit{flow} of the cluster state as it is cycled across the transition points. The dashed arrows indicate noise-induced barrier hopping events. \textbf{b}, (Top) Fluorescence images from the experiment while cycling the cluster across the transition via linear voltage ramps. The duration of each forward/backward ramp is 5 seconds. The octupole moments $\psi_{30}$ are computed from 3D reconstructed images using Abel transforms (see Methods). The solid lines show averaged trajectories over 90 such scans, with arrows marking the direction of the ramp. The cluster transitions from the pentagonal to a pyramidal arrangement sharply around $\alpha \!\approx\! 1.3$. The dot-dashed lines show a few backward sweeps where the pyramidal cluster changes abruptly to the pentagonal ground state over a wide range of $\alpha$. The region between the 16th and the 84th percentiles around the averaged trajectory is shaded. }
\end{figure*}

We have numerically computed the normal-mode frequencies of the square and the tetrahedral cluster
in the vicinity of the critical point $\alpha_c$ (see supplemental figure SF4); the spectrum of low-frequency modes is plotted as solid
lines in Fig.~\ref{fig:4ion-sm}(\textbf{c}). On both sides of the transition, we observe a zero-frequency mode corresponding to the free rotation about the trap axis and a mode that 
softens at $\alpha_c$. On the tetrahedral side, this zero mode and soft mode can be associated, respectively, with phase and amplitude fluctuations of the octupolar order parameter $\psi_{\rm 32}$, analogous to the Goldstone and the Higgs modes in the language of phase transitions.

Experimentally, we can couple to $|\psi_{\rm 32}|^2$ by modulating the trap \textit{rf} voltage. This modulation effectively drives the soft mode of the tetrahedron (see Methods) and leads to a blurring effect in the spacing between the upper and lower arms. Such coupling to the absolute value of the order parameter is an established route to exciting Higgs-like modes \cite{Podolsky2011_Higgs, Endres2012_Higgs}. The response of the 4-ion cluster in the fluorescence images is shown in Fig.~\ref{fig:4ion-sm}(\textbf{b}). Over a significant range of $\alpha$, we find that the peak response frequency is in excellent agreement with the numerically computed soft-mode spectrum, as seen from Fig.~\ref{fig:4ion-sm}(\textbf{c}).

We note that it is difficult to explore the regime very close to the transition point since it becomes increasingly harder 
to distinguish the perturbed and unperturbed images as the separation between the arms of the tetrahedron becomes smaller than the possible image resolution. We also cannot excite the out-of-plane soft mode of the square for $\alpha \!>\! \alpha_c$ with the trap voltage modulation, as the drive in this case couples
predominantly to a high-frequency in-plane breathing mode (see Methods).

\subsection{Hysteresis and a triple point}

The planar CME of the $5$-ion cluster for $\alpha\gg1$ is a regular pentagon in the \textit{x-y} plane. As $\alpha$ is reduced, the pentagon transitions to a square-base pyramid with the apex ion on the \textit{z-}axis, as shown in Fig.~\ref{fig:5ion-hys}. The lowest-order multipole that distinguishes these configurations is the real, parity-odd octupole moment $\psi_{3 0} \!\equiv\! \int d{\bf r} \rho_e({\bf r}) Y_{3,0}({\bf r})$, where $Y_{3,0}({\bf r}) \!\propto\! 5z^3-3r^2z$. This vanishes for the pentagon and takes on positive (negative) values for the pyramid with the apical ion
at $z>0$ ($z < 0$). Below, we discuss the complex energy landscape for this $5$-ion cluster, obtained numerically via the Nudged Elastic Band (NEB) technique \cite{Henkelman_NEB} (see Methods for further details), and correlate this with our experimental measurement of the hysteretic dynamics of $|\psi_{3 0}|$.

The global minimum shifts between the pentagon and the pyramid at $\alpha_t\approx1.328$ (see Supplementary discussion I), marked by a discontinuous jump in the normal-mode spectrum (see SF5). Surprisingly, we find that two out-of-plane vibrational modes of the pentagon also soften close to this transition point via a supercritical pitchfork bifurcation as shown in Fig.~\ref{fig:5ion-hys}(\textbf{a}). Analytically, the critical point ($\alpha_c$) of this buckling instability matches the \textit{reconstructive} transition point, $\alpha_c  \!\simeq\! \alpha_t$ (Supplementary discussion II). This feature is analogous to a triple point in thermodynamic phase transitions \cite{Jackson_triple_point}, where a \textit{displacive} and a \textit{reconstructive} transition occur simultaneously. In addition, the two degenerate, buckled pentagons survive as local minima over a small window, vanishing via saddle-node bifurcation at $\alpha \!\approx\! 1.29$ (Fig.~\ref{fig:5ion-hys}(\textbf{a})). By contrast, the pyramid remains stable throughout. This gives rise to a prominent hysteresis in the cluster configuration across the transition point $\alpha_t$.

\begin{figure*}[]
    \includegraphics[width=0.97\textwidth]{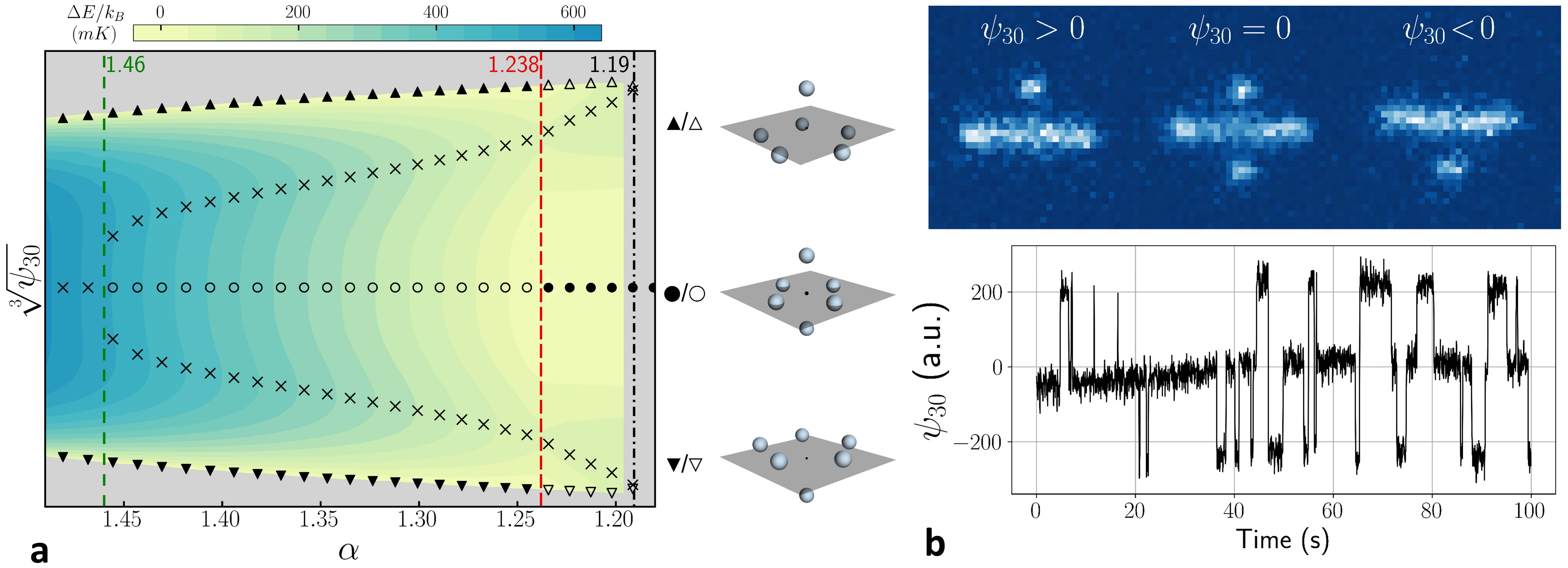}
    \caption{\label{fig:6ion-ss} \textbf{Stochastic switching between $6$-ion metastable configurations}. \textbf{a}, The energy landscape of the $6$-ion cluster around the pentagonal pyramid to a square bi-pyramid \textit{reconstructive} transition is shown as a function of $\alpha$ and $\sqrt[3]{\psi_{30}}$. The square bi-pyramid emerges as a stable node at $\alpha \!\approx\! 1.46$ via subcritical pitchfork bifurcation. The shift in the global minimum occurs at $\alpha\!\approx\!1.238$. The pyramids vanish from the landscape at $\alpha\!\approx\!1.19$ via saddle-node bifurcations. \textbf{b}, Stochastic jumps in the cluster state recorded close to the reconstructive transition $\alpha \!\approx\! 1.24$, where the forward and backward barrier heights are equal. The octupole moment $\psi_{30}$ of the cluster in each frame is computed using the Abel transform.}
\end{figure*}

By applying triangular voltage ramps for $U_{dc}$ (see Fig.~\ref{fig:fig1}), we drive the $5$-ion cluster back and forth in a series of axial compression and decompression cycles. In each ramp, as the axial confinement is weakened, the pentagon sharply transitions to a square pyramid beyond $\alpha_t$ with the apical ion in the pyramid randomly appearing at $z>0$ or $z<0$. This is signaled by the sudden increase in $|\psi_{30}|$ with decreasing $\alpha$ (rightward sweep) as shown in Fig.~\ref{fig:5ion-hys}(\textbf{b}). However, when the cluster is axially compressed by increasing $\alpha$ (leftward sweep), the pyramidal structure remains intact well beyond $\alpha_t$, as the system remains trapped in a local minimum. This phenomenon is analogous to the well-known example of metastable supercooled water. A photon kick or background collision triggers a nucleation-like event into the pentagonal ground state. These events occur over a broad range of $\alpha$, where the initial finite $|\psi_{30}|$ randomly jumps to zero. The resulting hysteresis loop is strongly asymmetric as seen from Fig.~\ref{fig:5ion-hys}(\textbf{b}). Real-time experimental videos demonstrating fast and slow nucleation events are provided as supplementary material.

\subsection{Stochastic switching}

In the $6$-ion cluster, a \textit{reconstructive} transition is observed between two 3D structures with different symmetries: A pentagonal pyramid with one apex ion on the $z$-axis switches to an octahedron (square bipyramid) with the two apex ions on the $z$-axis, as shown in Fig.~\ref{fig:6ion-ss}(\textbf{a}). The lowest-order multipole that distinguishes these configurations is again the real, parity-odd octupole moment $\psi_{3 0}$, which vanishes for the octahedron and takes 
on positive (negative) values for the pyramid with the apical ion
at $z>0$ ($z < 0$).

The energy landscape obtained via the NEB method is shown in Fig.~\ref{fig:6ion-ss}(\textbf{a}). The octahedron emerges as a stable node via subcritical pitchfork bifurcation at $\alpha\!\approx\!1.46$. The global minimum shifts between the pentagonal pyramid and the octahedron states at $\alpha\approx1.238$ (Supplementary discussion I), and is also marked by a discontinuous jump in the normal-mode spectra (see SF6). The two pyramids vanish from the potential landscape at $\alpha \!\approx\! 1.19$ via saddle-node bifurcations. The extended region of tri-stability between the octahedron and the two pentagon-pyramids leads to stochastic switching of the cluster configuration. 

We experimentally track the dynamics of $\psi_{3 0}$ close to the \textit{reconstructive} transition point, where all the minima have equal energy. 
Fig. \ref{fig:6ion-ss}(\textbf{b}) shows the time trace of $\psi_{30}$ obtained experimentally at $\alpha\!\approx\! 1.24$, which exhibits the expected three-state telegraph-like switching dynamics. We provide the video corresponding to the time trace as supplementary material.

\section{Discussion}

In summary, we investigate three distinct dynamical signatures of structural transitions in unit-cell-like 3D trapped-ion clusters. While our system lacks the ability to address individual ions, the diverse octupolar charge orders and rich energy landscapes that we present open avenues for further fundamental studies. By cooling the system to the motional ground state \cite{bollinger_groundstate_2019}, structural transitions can be studied in the quantum regime \cite{retzker_double_2008, shimshoni_quantum_2011, puebla_structural_2018, vuletic_aubrytransition_2021}. By introducing spin-spin interactions \cite{Blatt_sims, britton_engineered_2012, monroe-revmodphy} between ions, one can explore frustrated dynamics arising from competing charge and spin orders \cite{nath_hexagonal_2015, qiao_observing_2022, sun_two-dimensional_2024}|a frontier in many-body quantum physics. %

More readily, the stochastic switching between different cluster configurations can be applied to study reaction kinetics. Previously, such switching behavior was observed in \textit{zig-zag} ion chains in a regime where background gas collisions are the dominant mechanism \cite{Pagano_2019,aikyo_vacuum_characterization}. The ability to tune the barrier heights and noise sources (e.g., laser photon recoil) in 2D/3D energy landscapes \cite{mizukami_2025_isomerization} provides an ideal testbed for studies of thermal activation processes. One can also employ these controllable 2D and 3D clusters to investigate stochastic resonance \cite{yuan_stochastic_2024}, pattern formation \cite{pattern_formation_Lee}, heat transport and thermalization \cite{bermudez_controlling_2013, freitas_heat_2016, timm_heat_2023}, dynamics in the presence of impurities \cite{sias_melting} and topological defects \cite{del_campo_structural_2010, ulm_observation_2013, pyka_topological_2013, 2D_defect}, as well as Floquet engineering \cite{floquet_engineering_Kiefer}, broadening our understanding of collective nonlinear phenomena in the presence of long-range interactions.

\section{Materials and Methods}

\subsection{Experimental system}

The time-varying potential generated by the end-cap type trap shown in Fig. \ref{fig:fig1} is of the form:

\begin{figure}[b]
    \includegraphics[width=0.49\textwidth]{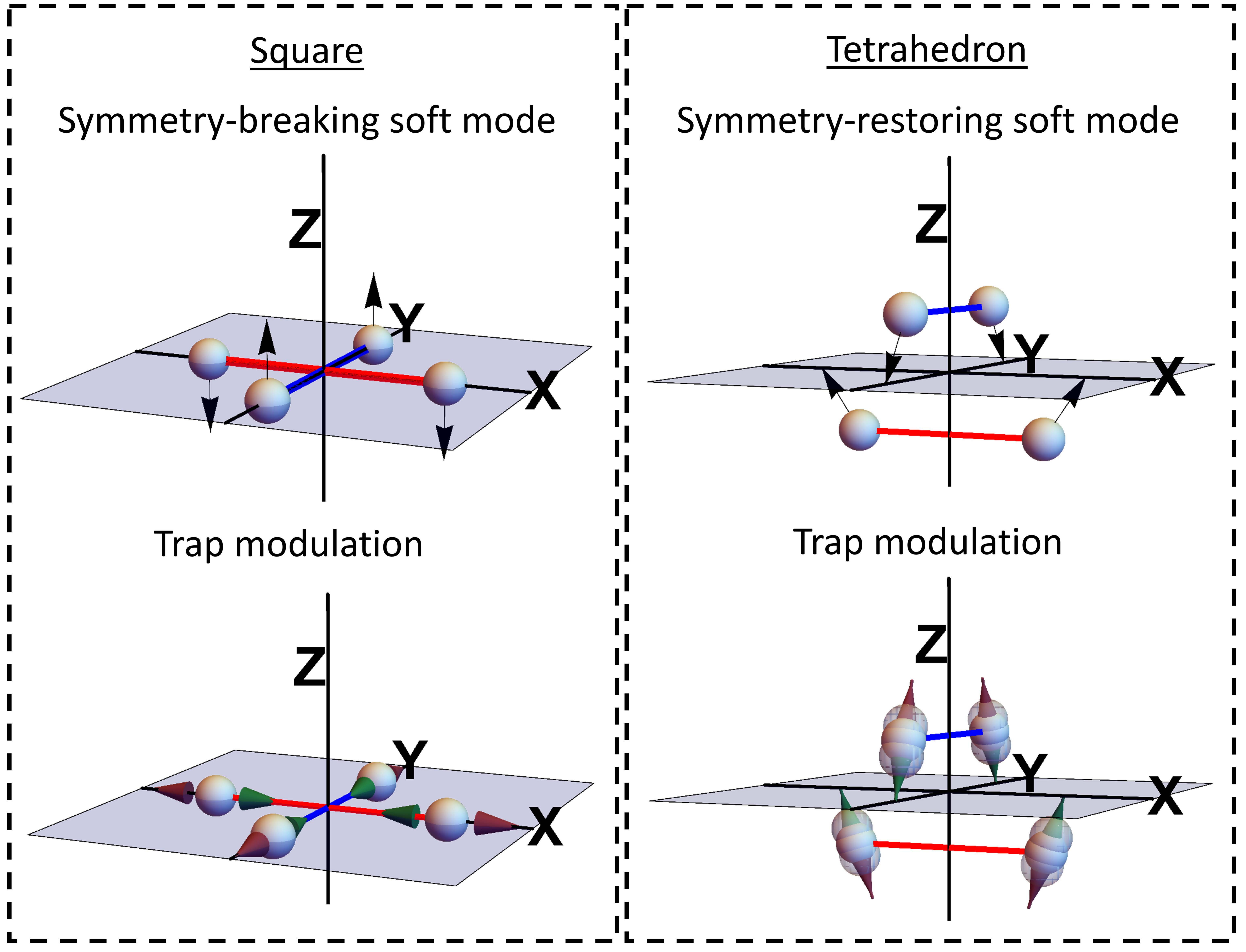}
    \caption{\label{fig:4i-sm-vectors} \textbf{(Top)} Soft-mode eigenvectors of square and tetrahedron. \textbf{(Bottom)} The arrows with the same color represent in-phase motion of each ion due to trap voltage modulation.}
\end{figure}

\begin{equation}
    \Phi(\boldsymbol{r}, t) = \eta (U_{dc} + V_{rf} \cos(\Omega_{rf} t)) \times\frac{x^2 + y^2 - 2z^2}{4 z_0^2},
\end{equation}
where $z_0$ is half the distance between \textit{rf}-electrodes in the axial direction and $\eta$ is the quadrupole efficiency of the trap geometry \cite{nisbet-jones_single-ion_2016,lindvall_high-accuracy_2022}.
The \textit{dc} ($U_{dc}$) and \textit{rf} ($V_{rf}$) voltages are set to operate the trap within the first Mathieu stability region \cite{werth_charged_2005}, represented in the rescaled Mathieu \textit{a} ($ \propto U_{dc}$) and \textit{q} ($ \propto V_{rf}$) parameter space. Under the adiabatic approximation \cite{leibfried_quantum_2003}, $|a|, q^2 \ll 1$, the angular frequencies of the time-averaged harmonic potential are related to the Mathieu parameters as:

\begin{eqnarray}
        \omega_{x,z}  & = & \frac{\Omega_{rf}}{2}\sqrt{\frac{q_{x,z}^2}{2}+a_{x,z}},
\end{eqnarray}
where $a_z = -2 a_x = \frac{4\eta QU_{dc}}{mz_0^2 \Omega_{rf}^2}$, $q_z =- 2 q_x = \frac{-2\eta QV_{rf}}{mz_0^2 \Omega_{rf}^2}$, \textit{Q} is the charge of the trapped particles ($Q=+e$ for singly ionized Calcium), and $\Omega_{rf}=18.26$ MHz is the drive frequency of the trap. By fixing $V_{rf} \sim 235$ V and varying $U_{dc} \sim 1$ V to $-12$ V, $\alpha=\omega_z / \omega_x$ can be continuously tuned from $\sim 2.2$ to $0.25$. The fluorescence images of the clusters are captured onto an EMCCD camera with a custom-made objective of $\times$22 magnification.

\subsection{Trapped ion motional excitation}

We excite the motion of trapped ions in the experiment via two different methods. The first method is forced oscillation of the trap center, where one of the conical DC electrodes is given a small sinusoidal voltage, which results in the center-of-mass (c.o.m.) of the cluster oscillating about the trap center. When this frequency matches the c.o.m. mode's frequency, the increase in the oscillation amplitude of the ions is seen on EMCCD. Any slight misalignment of the conical electrode that is inadvertent leads to an oscillatory gradient field along both the radial and axial directions, enabling the excitation of both c.o.m. modes on resonance with this scheme. We directly measure $\alpha$ at different $U_{dc}$ values by exciting a single ion's radial and axial motion. The second method is the amplitude modulation of \textit{rf}-voltage ($V_{rf}$) at a frequency $\Omega_{m}$. This leads to an oscillatory Mathieu $q$ parameter, effectively modulating the elastic constants of the time-averaged harmonic potential, $\omega_x^2$ and $\omega_z^2$. This provides an alternative scheme for parametric excitation of the c.o.m.~motion for $\Omega_{m} = 2 \times \omega_{x,z}$ \cite{schmidt_ion_removal}.

\begin{table}[t]
\caption{\label{tab:table1}%
\textbf{Projections of the driven motion due to voltage modulation onto the 12 normal-mode eigenvectors of the $4$-ion cluster.}
}
\begin{ruledtabular}
\begin{tabular}{ccccc}
\textrm{Normal-mode} & \multicolumn{2}{c}{\textrm{Square}} & \multicolumn{2}{c}{\textrm{Tetrahedron}} \\
& \multicolumn{2}{c}{($\alpha = 1.254$)}  & \multicolumn{2}{c}{($\alpha = 1.186$)}  \\
& $\omega_k / \omega_x$ & $| \bm{v}_{m} \cdot \bm{u}_k |$  & $\omega_k / \omega_x$ & $| \bm{v}_{m} \cdot \bm{u}_k |$  \\
\colrule
Vibrational-mode 1 & 1.73 & 1 & 1.76 & 0.09 \\
Vibrational-mode 2 & 1.49 & 0 & 1.37 & 0 \\
Vibrational-mode 3 & 1.32 & 0 & 1.35 & 0 \\
Vibrational-mode 4 & 1.32 & 0 & 1.35 & 0 \\
c.o.m. - Z & 1.254 & 0 & 1.186 & 0 \\
c.o.m. - X & 1    & 0 & 1    & 0 \\
c.o.m. - Y & 1    & 0 & 1    & 0 \\
Vibrational-mode 5 & 0.88 & 0 & 0.94 & 0 \\
X - rotation & 0.76 & 0 & 0.52 & 0 \\
Y - rotation & 0.76 & 0 & 0.52 & 0 \\
\textbf{soft mode} & \textbf{0.3}  & \textbf{0} & \textbf{0.37} & \textbf{0.996} \\
Z - rotation & 0    & 0 & 0    & 0 \\
\end{tabular}
\end{ruledtabular}
\end{table}

However, in the case of a multi-ion cluster, the modulation of trap curvatures will also drive the ions about the equilibrium configuration $\bm{v}_0$ along the unit vector $\bm{v}_{m}$. We obtain this unit vector at a set of ($V_{rf}$, $U_{dc}$) values, by numerically computing the CME coordinates $\bm{v}_0^{\pm \delta}$ corresponding to small perturbation in \textit{rf}-voltage $V_{rf}(1 \pm \delta)$. For small values of $\delta$ $(\sim 10^{-3})$,

\begin{equation}
    \bm{v}_{m} = \frac{\bm{v}_0^{+ \delta} - \bm{v}_0^{- \delta}}{||\bm{v}_0^{+ \delta} - \bm{v}_0^{- \delta}||}.
\end{equation}
Collective normal modes (eigen vectors $\bm{u}_k$) that can couple to this drive have a finite projection $| \bm{v}_{m} \cdot \bm{u}_k |$.

Let us look at the effect of trap voltage modulation on the collective modes of $4$-ion cluster near the square to tetrahedron transition. The projections between the drive and the normal modes are tabulated in Table 1. The symmetry-restoring soft mode of the tetrahedron (shown in Fig.~\ref{fig:4i-sm-vectors}) has a projection $\sim1$, enabling the resonant excitation of this mode experimentally. The same technique, however, excites a high-frequency in-plane mode of the square shown in Fig.~\ref{fig:4i-sm-vectors}, and cannot be used to observe the softening of the square's symmetry-breaking mode. To demonstrate the resonant normal mode excitation via curvature modulation, we present a simple toy problem of two ions in a one-dimensional harmonic potential in the supplementary discussion III. In this simple case, the modulation of the spring constant leads to an on-resonant excitation of the breathing mode.

\begin{figure}[t]
    \includegraphics[width=0.49\textwidth]{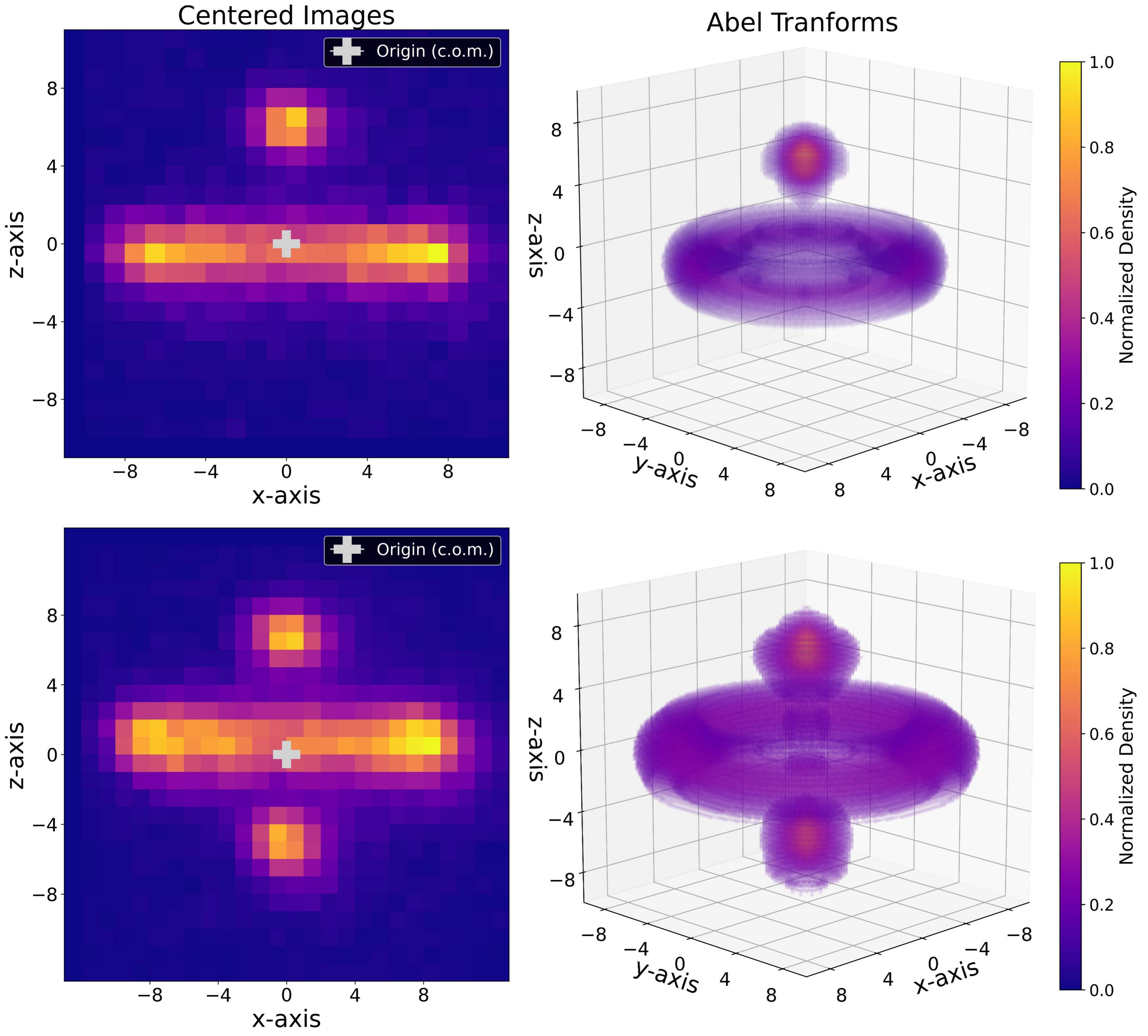}
    \caption{\label{fig:abel-transforms} Heat maps showing Abel transforms of the $6$-ion pyramid (top) and the octahedron (bottom) obtained from fluorescence images to calculate the octupole moment of the charge distribution.}
\end{figure}

\subsection{Numerical computations of the energy landscape}

The equilibrium positions of the ions can be efficiently obtained by minimizing the total potential energy (eq.~1) about a guess configuration, using the built-in \textit{FindMinimum} function of Mathematica. For each $\alpha$ value, the initial configuration of the cluster is randomized and then steered to a local minimum. This is repeated several times to identify the global minimum (CME). The collective-mode eigenvectors ($\bm{u}_k$) and their corresponding frequencies ($\omega_k$) at the theoretical limit $T=0$ are then obtained from the Hessian matrix
\begin{equation}
    H_{ij} = \left. \frac{\partial^2 V}{\partial v_{i} \partial v_{j}} \right|_{\bm{v}_0}  ,
\end{equation}
where $v_i$ are the components of the vector $\bm{v}=(x_1, y_1, z_1,..., x_n, y_n, z_n)$ and $\bm{v}_0$ is the coordinate of the equilibrium configuration (local or global minimum).

To obtain the minimum energy path (MEP) and the saddle point between two equilibrium configurations, we employ a chain of states method. A guess path between the stable equilibria is constructed using linear interpolation. The entire chain is taken to the MEP with the NEB method \cite{Henkelman_NEB}. Around the configuration with maximum energy in the MEP, we find an unstable equilibrium in the potential energy landscape, with one negative eigenvalue of its corresponding Hessian. The contour plots of energy landscapes are obtained by stitching the MEPs between consecutive minima, sorted by a suitable order parameter.

\subsection{Computing moments from images}

The presence of cylindrical symmetry about \textit{z-}axis in our system permits us to apply Abel transforms \cite{abel_1826, hickstein_direct_2019} on the fluorescence images to reconstruct a time-averaged 3D charge distribution. Fig.~\ref{fig:abel-transforms} shows the Abel transforms obtained for the $6$-ion 3D orientations|pentagonal pyramid and octahedron. We use this 3D charge distribution to compute the octupole moments $\psi_{30}$ for the $5$-ion hysteretic dynamics, and the $6$-ion stochastic switching.

\section*{Acknowledgements}
We thank Arun Roy for comments on the manuscript. The authors acknowledge support from the Department of Science and Technology and the Ministry of Electronics and Information Technology (MeitY), Government of India, under the “Centre for Excellence in Quantum Technologies” grant with Ref. No. 4(7)/2020-ITEA. A. Paramekanti acknowledges funding from the Natural Sciences and Engineering Research Council (NSERC) of Canada.

%




\onecolumngrid
\clearpage

\begingroup

\setstretch{1.2}
\setcounter{figure}{0}
\setcounter{section}{0}

\renewcommand{\thefigure}{SF\arabic{figure}}

\setcounter{equation}{0}
\renewcommand{\theequation}{S\arabic{equation}}

\setcounter{page}{1}
\renewcommand{\thepage}{\arabic{page}}

\begin{center}
{\LARGE{\textbf{Supplementary discussion}}}
\end{center}

\section{Analytical calculations of transition points}

For the ease of calculating the transition points, the energy and length can be scaled into dimensionless units as $V \rightarrow V/E_{\perp}$ and $\bm{r} \rightarrow \bm{r}/l_\perp$, where
\begin{equation}
    l_\perp =  \left(\frac{k_ee^2}{m\omega_x^2} \right)^{1/3} \quad \text{and} \quad
    E_\perp =  \left( m\omega_x^2 k_e^2e^4 \right)^{1/3}.
\end{equation}

\begin{figure}[b]
    \includegraphics[width=0.99\textwidth]{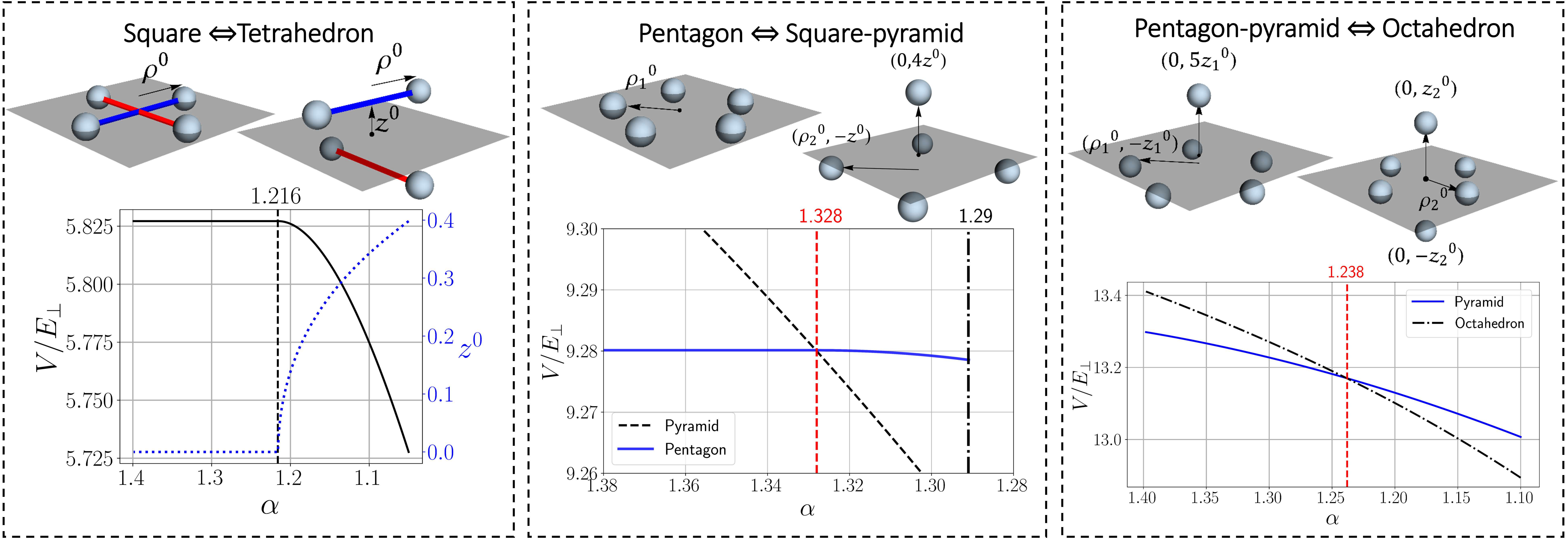}
    \caption{\label{fig:transition_points} The CMEs of 4-, 5-, and 6-ion clusters at the transition points of interest. }
\end{figure}

\subsection{4-ion square to tetrahedron}

In these scaled units, the potential energy of the tetrahedral configuration in terms of the coordinates $\rho$, $z$ (depicted in Fig. \ref{fig:transition_points}) is
\begin{equation}
    V_{4}(\rho,z) = 2\rho^2 + 2 \alpha^2 z^2 + \frac{1}{\rho} + \frac{4}{\sqrt{2\rho^2 + 4z^2}},
\end{equation}

Simultaneously solving $\partial V_{4}/\partial\rho = 0$ and $\partial V_{4}/\partial z = 0$ for the equilibrium positions $(\rho^0 ,z^0)$, we obtain
\begin{equation}
    z^0 = \begin{cases} 
          0 & \alpha\geq \alpha_c \\
          \frac{1}{2} \,\left[\frac{1}{4}\left( \frac{2}{\alpha^2} \right)^{2/3} - \frac{1}{2} \left( \frac{1}{4(2-\alpha^2)} \right)^{2/3} \right]^{1/2} & \alpha < \alpha_c
       \end{cases}
\end{equation}
where $\alpha_c = 2\sqrt{\frac{4-\sqrt{2}}{7}}\approx1.216$ is the critical point where $z^0$ becomes non-zero, breaking inversion symmetry.

\subsection{5-ion pentagon to square-pyramid}

The potential energies in scaled units of the pentagonal and the square-pyramidal configurations are, respectively,

\begin{align}
    V^{pentagon}_5(\rho_1) =&\;  \frac{5}{2}\rho_1^2+\frac{\sqrt{25+10\sqrt{5}}}{\rho_1},\\
    V^{pyramid}_5(\rho_2,z) =&\;  2\rho_2^2+10\alpha^2 z^2+\frac{1+2\sqrt{2}}{\rho_2}+\frac{4}{\sqrt{\rho_2^2+25z^2}},
\end{align}
where the coordinates $\rho_1$, $\rho_2$, and $z$ are depicted in Fig. \ref{fig:transition_points}. Solving for the equilibrium configurations $\rho_1^0$ and $(\rho_2^0, z^0)$, respectively, the minimum energy of the pentagon is obtained as $V^{pentagon}_5(\rho_1^0) \approx 9.28$, while the minimum energy of the square-pyramid decreases with $\alpha$. Both curves intersect at $\alpha_t\approx 1.328$ as shown in Fig. \ref{fig:transition_points}.

\subsection{6-ion pentagonal-pyramid to octahedron}

The potential energies in scaled units of the pentagonal-pyramid and the octahedron configurations are, respectively,

\begin{align}
    V^{pyramid}_6(\rho_1,z_1) =&\;  \frac{5}{2}\rho_1^2+15 \alpha^2 z^2_1+\frac{\sqrt{25+10\sqrt{5}}}{\rho_1}+\frac{5}{\sqrt{\rho^2_1 + 36z^2_1}},\\
    V^{octahedron}_6(\rho_2,z_2) =&\;  2\rho_2^2+\alpha^2 z^2_2+\frac{1+2\sqrt{2}}{\rho_2}+\frac{8}{\sqrt{\rho_2^2+z^2_2}}+\frac{1}{2z_2}.
\end{align}
where the coordinates ($\rho_1$, $z_1$) and ($\rho_2$, $z_2$) are depicted in Fig. \ref{fig:transition_points}. The minima of these functions intersect at $\alpha\approx1.238$.

\section{Mode softening of the 5-ion pentagon}

\subsection{Out-of-plane normal modes}

In the pentagonal configuration, the ion positions are given by $\mathbf{r}_n = \left( \rho \cos\left(\frac{2\pi n}{5}\right), \rho \sin\left(\frac{2\pi n}{5}\right), 0 \right)$, where $\rho$ is the circum-radius of the pentagon. The total potential energy of the pentagonal configuration is:
\[
V(\rho) = \frac{5}{2} \rho^2 + \frac{S_5}{\rho}
\]
where $S_5 = \sqrt{25 + 10 \sqrt{5}}$. The equilibrium radius $\rho_0$ is found by minimizing $V(\rho)$ with respect to $\rho$, which gives

\[
\rho_0^3 = \frac{S_5}{5} \;.
\]

\begin{figure}
    \centering
    \includegraphics[width=0.95\textwidth]{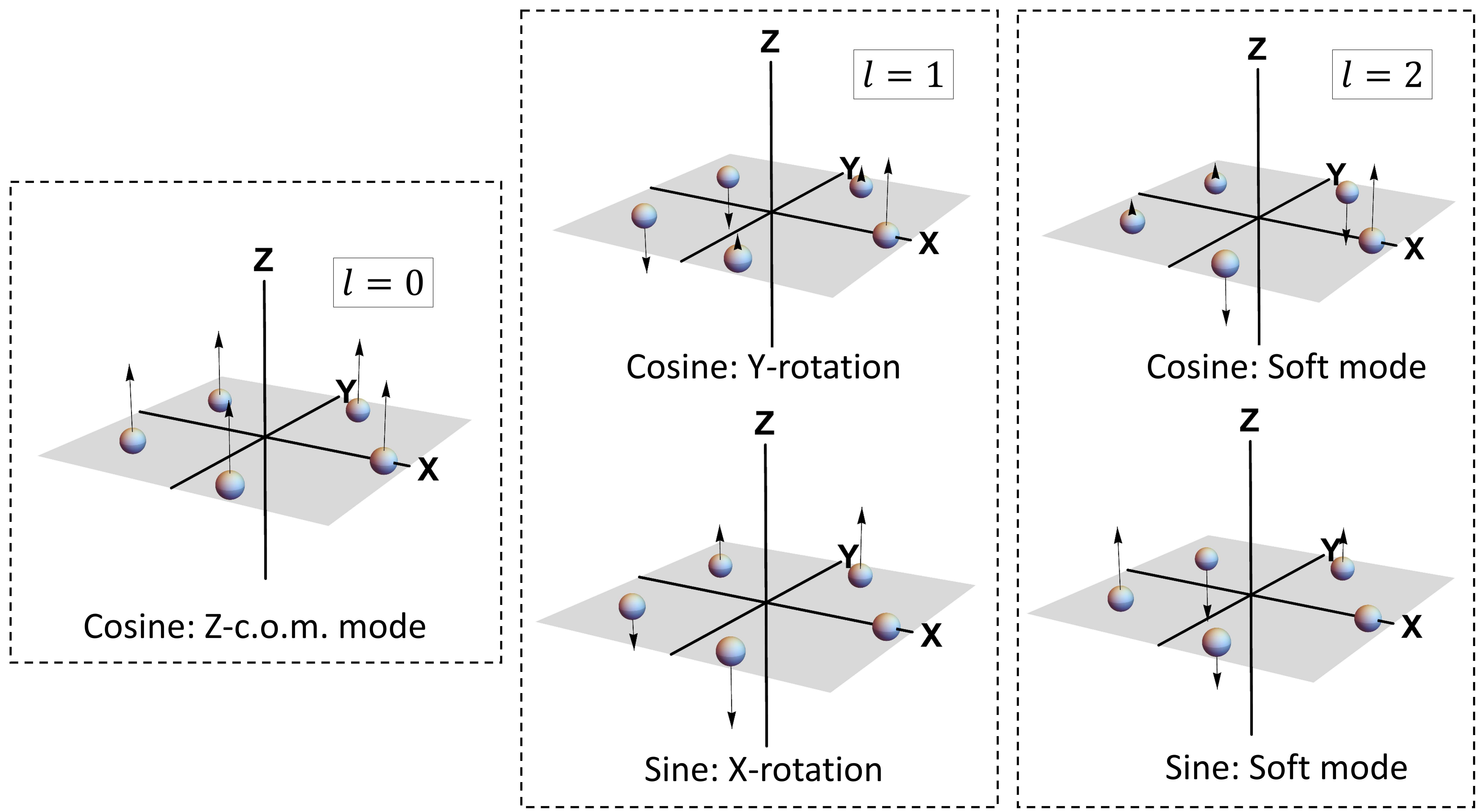}
    \caption{The out-of-plane collective oscillations of the 5-ion planar pentagon.}
    \label{fig:pentagon-modes}
\end{figure}

The $z$-displacements of the ions for the fundamental out-of-plane modes \cite{podolsky_buckling} can be described by sinusoidal functions of the ion index \textit{n} (where $n=0, 1, 2, 3, 4$):
\begin{equation}
    z_n \propto 
    \begin{cases}
        \cos\left(\frac{2\pi l n}{5}\right) & l=0,1,2 \\[5pt]
        \sin\left(\frac{2\pi l n}{5}\right) & l=1,2
    \end{cases}
\end{equation}

where $l$ is an integer characterizing the angular dependence. These modes are sketched in Fig. \ref{fig:pentagon-modes}: c.o.m. Z-Motion for Cosine \textit{l}=0 mode, rigid tilt/rotation about Y-axis for Cosine \textit{l}=1 mode, rigid tilt/rotation about X-axis for Sine \textit{l}=1 mode, and two buckling modes for Cosine and Sine \textit{l}=2.

\subsection{Buckling critical point}

The ion positions are slightly perturbed along the sine buckling mode:
\[
\mathbf{r}_n^0(\varepsilon) = \left( \rho_0 \cos\left(\frac{2\pi n}{5}\right), \rho_0 \sin\left(\frac{2\pi n}{5}\right), \varepsilon \sin\left(\frac{4\pi n}{5}\right) \right)
\]
We now calculate the change in total potential energy $\Delta V = V(\rho_0, \varepsilon) - V(\rho_0)$ to order $\varepsilon^2$. The change in harmonic potential energy is positive, given by
\[
\Delta V_{\text{trap}}^{(s)} = \frac{1}{2} \alpha^2 \varepsilon^2 \sum_{n=0}^4 \sin^2\left(\frac{2\pi n}{5}\right) \;.
\]

The distance between ions $i$ and $j$ is $|\mathbf{r}_i - \mathbf{r}_j| = \sqrt{(d_{ij}^0)^2 + (z_i^{(c)} - z_j^{(c)})^2}$, where the in-plane distance at the equilibrium configuration is $d_{ij}^0 = |\mathbf{r}_i^0 - \mathbf{r}_j^0| = 2\rho^0 \left|\sin(\frac{\pi(i-j)}{5})\right|$. For small $\varepsilon$, we expand the inverse distance:
\begin{align}
\frac{1}{|\mathbf{r}_i - \mathbf{r}_j|} &= \frac{1}{d_{ij}^0} \left( 1 + \frac{(z_i^{(c)} - z_j^{(c)})^2}{(d_{ij}^0)^2} \right)^{-1/2}\\ 
&\approx \frac{1}{d_{ij}^0} \left( 1 - \frac{(z_i^{(c)} - z_j^{(c)})^2}{2 (d_{ij}^0)^2} \right) \;.
\end{align}
The change in Coulomb energy is
\begin{align*}
\Delta V_{\text{Coulomb}}^{(s)} &=\sum_{i<j}\left(\frac{1}{|\mathbf{r}_i - \mathbf{r}_j|} - \frac{1}{d_{ij}^0}\right)\\ 
&\approx \sum_{i<j} \left( - \frac{(z_i^{(s)} - z_j^{(s)})^2}{2 (d_{ij}^0)^3} \right) \\
&= - \frac{ \varepsilon^2}{2} \sum_{i<j} \frac{\left[ \sin\left(\frac{4\pi i}{5}\right) - \sin\left(\frac{4\pi j}{5}\right) \right]^2}{(d_{ij}^0)^3} \\
&\approx - \frac{ \varepsilon^2}{16 \rho_0^3} \sum_{i<j} \frac{\left[ \sin\left(\frac{4\pi i}{5}\right) - \sin\left(\frac{4\pi j}{5}\right) \right]^2}{|\sin(\pi(i-j)/5)|^3} \;.
\end{align*}
Let us define the dimensionless structural sum as $S_s$.

So the change in Coulomb energy is
\begin{equation}
\Delta V_{\text{Coulomb}}^{(s)} \approx - \frac{S_s}{16 \rho_0^3} \varepsilon^2 \;.
\label{eq:dV_coul_s}
\end{equation}

For the cosine $l=2$ mode, a similar analysis can be performed, which leads to the dimensionless structural sum in the Coulomb energy term:

\begin{equation}
S_c = \sum_{i<j, \, i,j=0..4} \frac{\left[ \cos\left(\frac{4\pi i}{5}\right) - \cos\left(\frac{4\pi j}{5}\right) \right]^2}{|\sin(\pi(i-j)/5)|^3} \;.
\end{equation}

\textbf{To show $S_c = S_s$}

\begin{align*}
S_c &= \sum_{i<j} \frac{\left[ \cos\left(\frac{4\pi i}{5}\right) - \cos\left(\frac{4\pi j}{5}\right) \right]^2}{|\sin(\pi(i-j)/5)|^3} \\
&= \sum_{i<j} \frac{4 \sin^2\left(\frac{2\pi(i+j)}{5}\right) \sin^2\left(\frac{2\pi(i-j)}{5}\right)}{|\sin(\pi(i-j)/5)|^3} \quad 
\\
&= \sum_{k=1,2} \left( \frac{4 \sin^2(2\pi k/5)}{|\sin(\pi k/5)|^3} \sum_{\substack{i<j \\ |i-j|\equiv k}} \sin^2\left(\frac{2\pi(i+j)}{5}\right) \right) \;. \quad 
\end{align*}

\begin{align*}
S_s&= \sum_{i<j} \frac{\left[ \sin\left(\frac{4\pi i}{5}\right) - \sin\left(\frac{4\pi j}{5}\right) \right]^2}{|\sin(\pi(i-j)/5)|^3}  
\\
&= \sum_{i<j} \frac{4 \cos^2\left(\frac{2\pi(i+j)}{5}\right) \sin^2\left(\frac{2\pi(i-j)}{5}\right)}{|\sin(\pi(i-j)/5)|^3} \quad
\\
&=  \sum_{k=1,2} \left( \frac{4 \sin^2(2\pi k/5)}{|\sin(\pi k/5)|^3} \sum_{\substack{i<j \\ |i-j|\equiv k}} \cos^2\left(\frac{2\pi(i+j)}{5}\right) \right) \;. \quad
\end{align*}

\begin{align*}
 S_s - S_c &= \sum_{i<j} \frac{4 \sin^2\tfrac{2\pi(i-j)}{5}}{|\sin(\pi(i-j)/5)|^3} \left( \cos^2\tfrac{2\pi(i+j)}{5} - \sin^2\tfrac{2\pi(i+j)}{5} \right) \\
 &= \sum_{k=1,2} \frac{4 \sin^2(2\pi k/5)}{|\sin(\pi k/5)|^3} \left( \sum_{\substack{i<j \\ j-i = k}} \cos\left(\frac{4\pi(i+j)}{5}\right) \right) \\
 &= \sum_{k=1,2} \frac{4 \sin^2(2\pi k/5)}{|\sin(\pi k/5)|^3} \times (0) \quad \text{(since } \sum_{m=0}^{4} \cos(4\pi m / 5) = 0 \text{)} \\
 &= 0 \;.
\end{align*}

Since $\sum_{n=0}^4 \sin^2\left(\frac{2\pi n}{5}\right) = \sum_{n=0}^4 \cos^2\left(\frac{2\pi n}{5}\right) = \frac{5}{2}$, the total change in potential energy to order $\varepsilon^2$ for both modes is
\begin{align*}
\Delta V_{\text{total}}^{(c)} &= \Delta V_{\text{total}}^{(s)} \\
&=\Delta V_{\text{trap}}^{(s)} + \Delta V_{\text{Coulomb}}^{(s)}\\
&= \left( \frac{5}{4} \alpha^2  - \frac{S_s}{16 \rho_0^3} \right) \varepsilon^2 \;.
\end{align*}

The critical point $\alpha = \alpha_c$ occurs when the coefficient is zero:
\[
\frac{5}{4} (\alpha_c)^2 - \frac{S_s }{16 \rho_0^3} = 0 \;.
\]
Solving for $(\alpha_c)^2$ and substituting $\rho_0^3 = S_5 / 5$:
\[
(\alpha_c)^2 = \frac{S_s}{20  \rho_0^3} = \frac{S_s}{4 S_5}
\]
We find:
\begin{equation}
\boxed{\alpha_c = \sqrt{\frac{S_s}{4 S_5}} \approx 1.328 \;.}
\label{eq:alpha_c_prime_sq}
\end{equation}

\newpage

\begin{figure}[t]
    \includegraphics[width=0.99\textwidth]{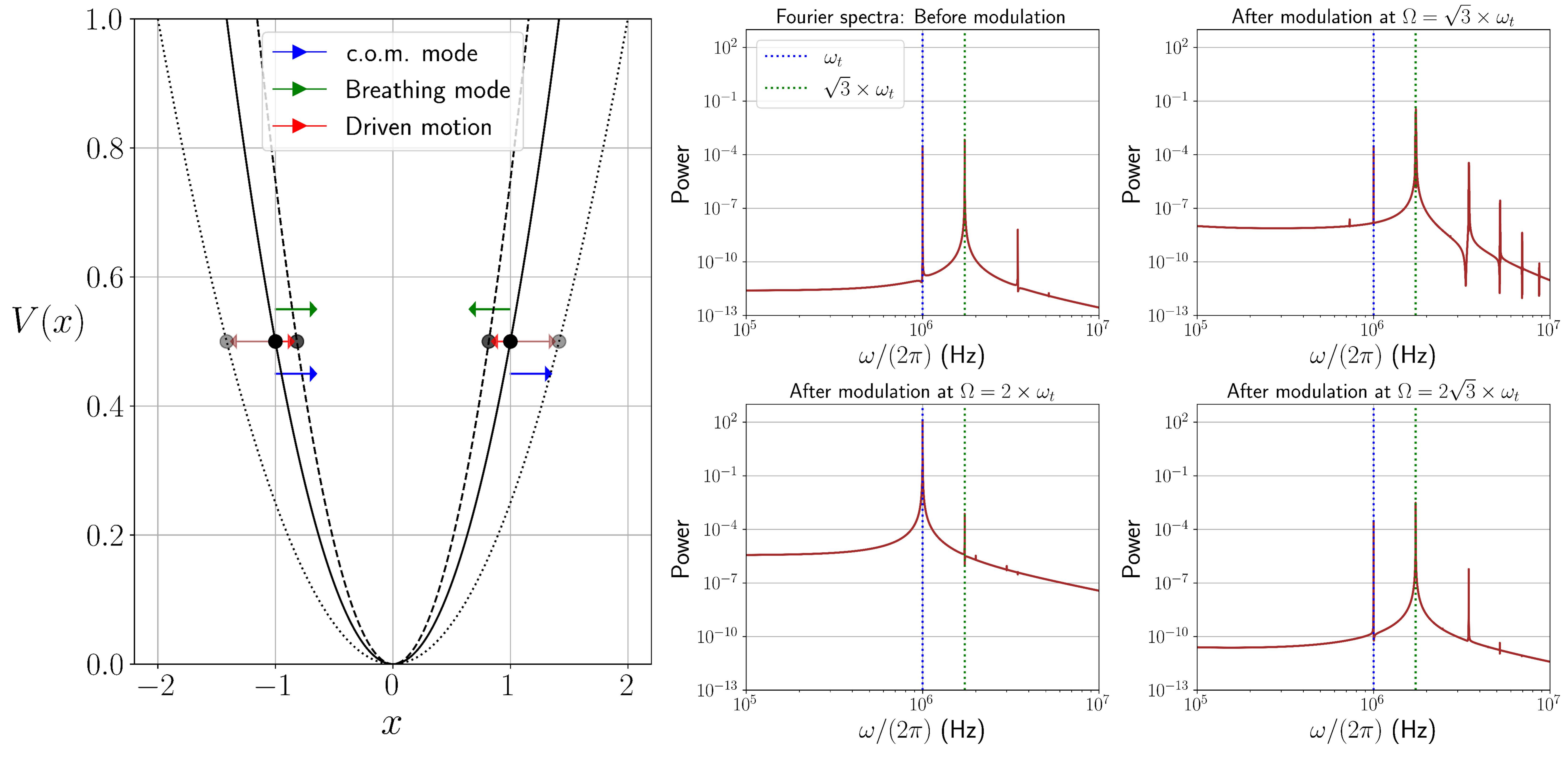}
    \caption{\label{fig:2ion-mod} Toy-problem of 2 ions trapped in 1D harmonic potential. The two normal modes are c.o.m. oscillation (blue arrows) and breathing mode (green arrows). Modulation of the trap curvature leads to the driven motion along the red arrows. The Fourier spectrum of the motion shows on-resonant excitation of the breathing mode via curvature modulation. }
\end{figure}

\section{Collective mode excitation by modulating spring constant}

Consider a toy problem of two like charges with mass \textit{m}, co-trapped in a 1D harmonic potential of frequency $\omega_{t}=\sqrt{2k/m} = 2 \pi \times 1$ MHz as shown in Fig. \ref{fig:2ion-mod}. The two normal modes are: c.o.m. mode with frequency $\omega_{c.o.m.} = \omega_{t}$ and breathing mode with frequency $\omega_{b} = \sqrt{3} \times \omega_{t}$. The motion of the charges is first damped to equilibrium positions, and the spring constant $k$ is then modulated at $\Omega_{m}$, i.e., $k\rightarrow k(1+\epsilon \cos(\Omega_{m}t))$. In Fig. \ref{fig:2ion-mod}, the power spectrum of the ions' motion after the damping period is compared with the power spectra after the modulation at different frequencies. For a modulation strength $\epsilon=10^{-3}$, the c.o.m. mode shows parametric excitation at $\Omega_{m} = 2 \times \omega_t$, while the breathing mode shows resonant excitation for $\Omega_m = \sqrt{3} \times \omega_t$. In this case, it is trivial to see that the unit vector along the driven motion overlaps is along the breathing mode (Fig. \ref{fig:2ion-mod}).

\newpage

\textbf{Supplementary video 1: \texttt{five\_ions\_hysteresis.mp4}}

\vspace{5pt}

This is a real-time video of the 5-ion cluster, capturing the full range of hysteresis in a single voltage ramp. During the upward ramp, the apex pyramidal ion pops out of the plane at $\alpha\sim1.3$. During the downward ramp, it goes back into the plane at a much higher value of $\alpha\sim1.56$. 

\vspace{5pt}

\textbf{Supplementary video 2: \texttt{five\_ions\_nucleation.mp4}}

\vspace{5pt}

This is a real-time video of the 5-ion cluster, capturing a fast nucleation event from the metastable pyramid to the stable pentagon. During the upward ramp, the apex pyramidal ion pops out of the plane at $\alpha\sim1.3$. During the downward ramp, it goes back into the plane at $\alpha\sim1.38$. 

\vspace{5pt}

\textbf{Supplementary video 3: \texttt{six\_ions\_switch.mp4}}

\vspace{5pt}

A video of the 6-ion cluster switching between pyramidal and bi-pyramidal configurations. The plot in Fig. 4 is obtained from this video.

\newpage

\begin{figure*}[]
    \includegraphics[width=0.99\textwidth]{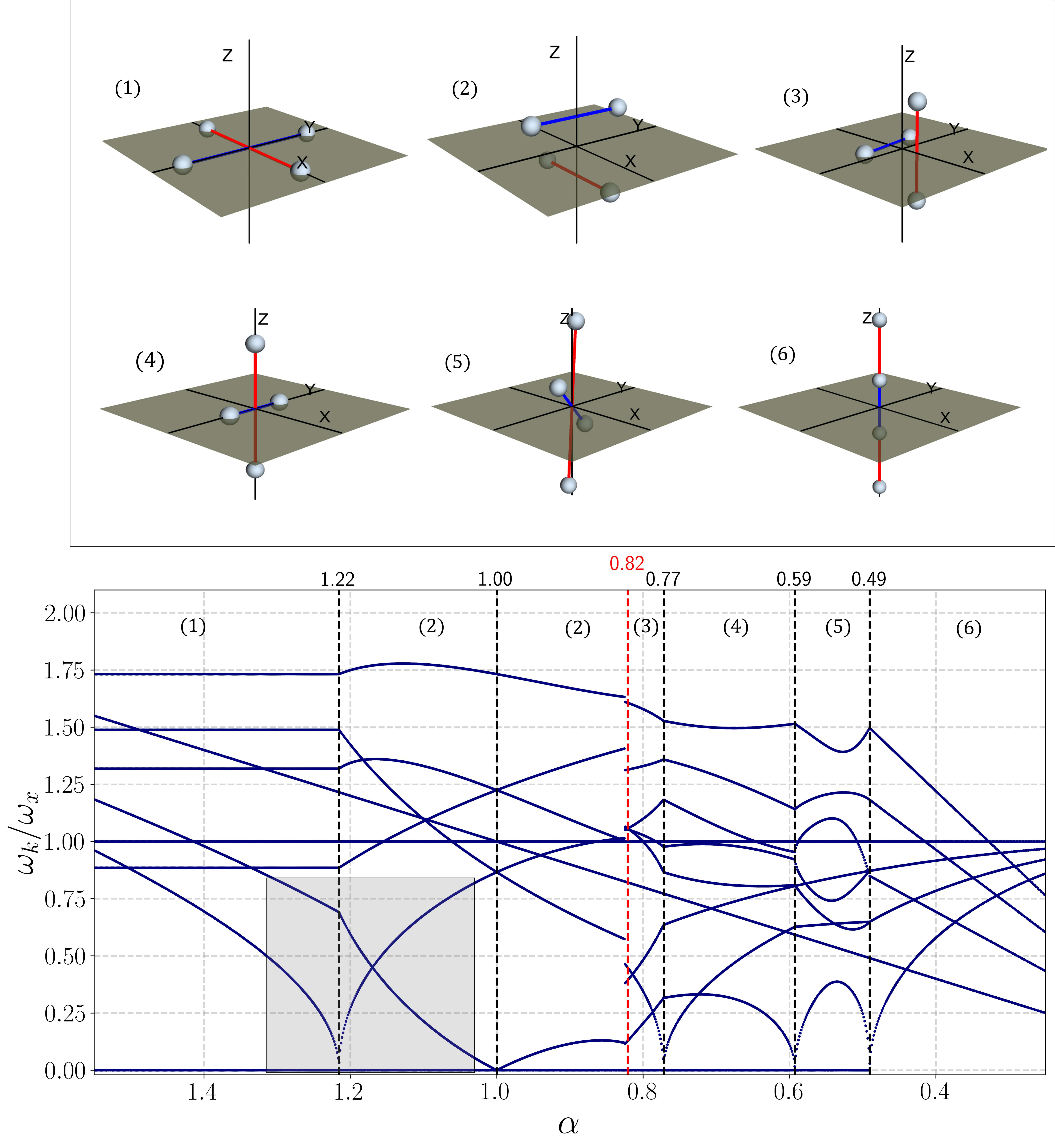}
    \caption{\label{fig:4ion-cmenm} \textbf{(Top)} All the 4-ion minimum energy configurations. \textbf{(Bottom)} Corresponding normal-mode frequencies computed at the theoretical limit $T=0$. The grey region is the ROI for the study of Higgs-like mode softening at a symmetry-breaking transition.}
\end{figure*}

\begin{figure*}[]
    \includegraphics[width=0.99\textwidth]{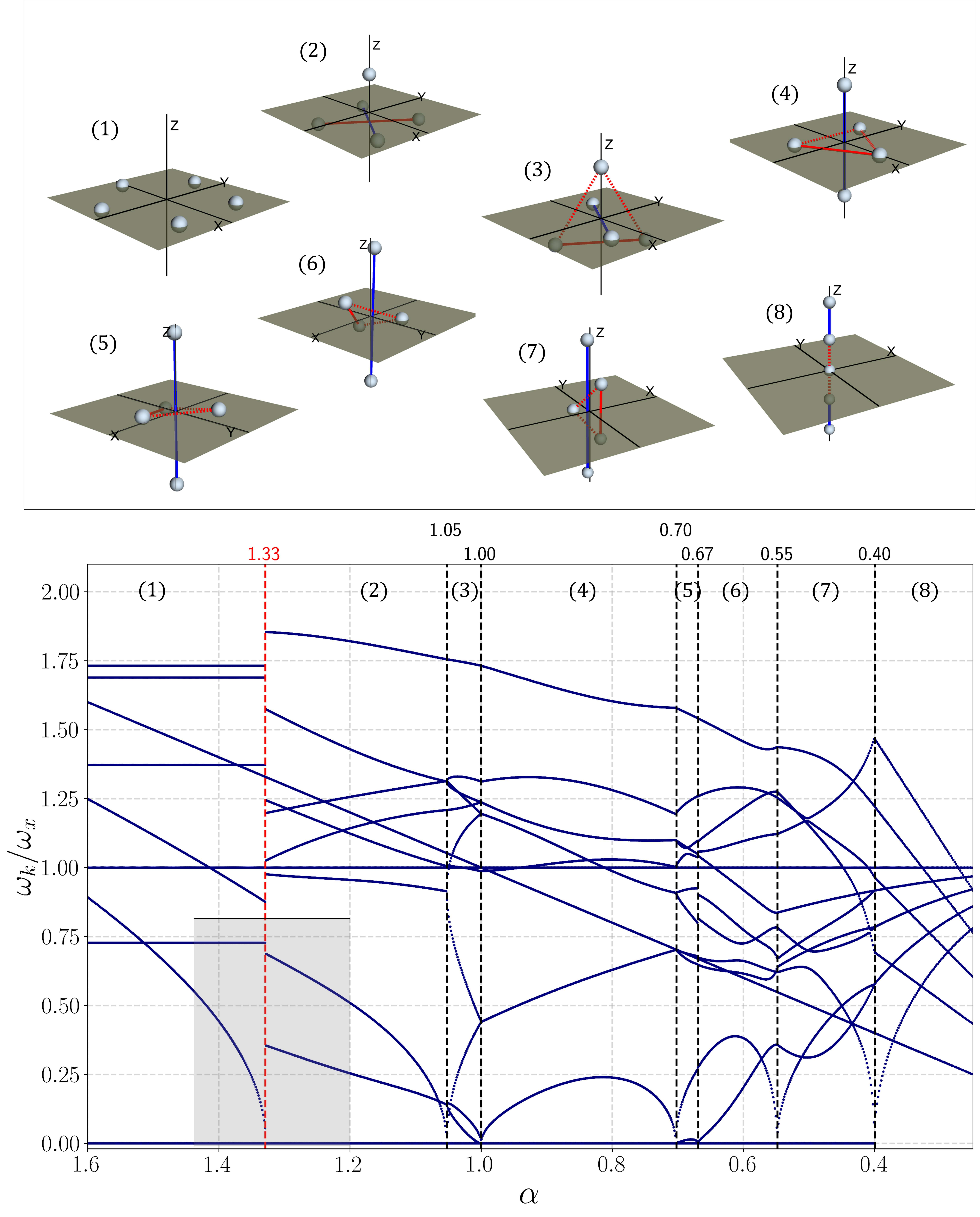}
    \caption{\label{fig:5ion-cmenm} \textbf{(Top)} All the 5-ion minimum energy configurations. \textbf{(Bottom)} Corresponding normal-mode frequencies computed at the theoretical limit $T=0$. The grey region is the ROI for the study of hysteresis and a triple point.}
\end{figure*}

\begin{figure*}[]
    \includegraphics[width=0.99\textwidth]{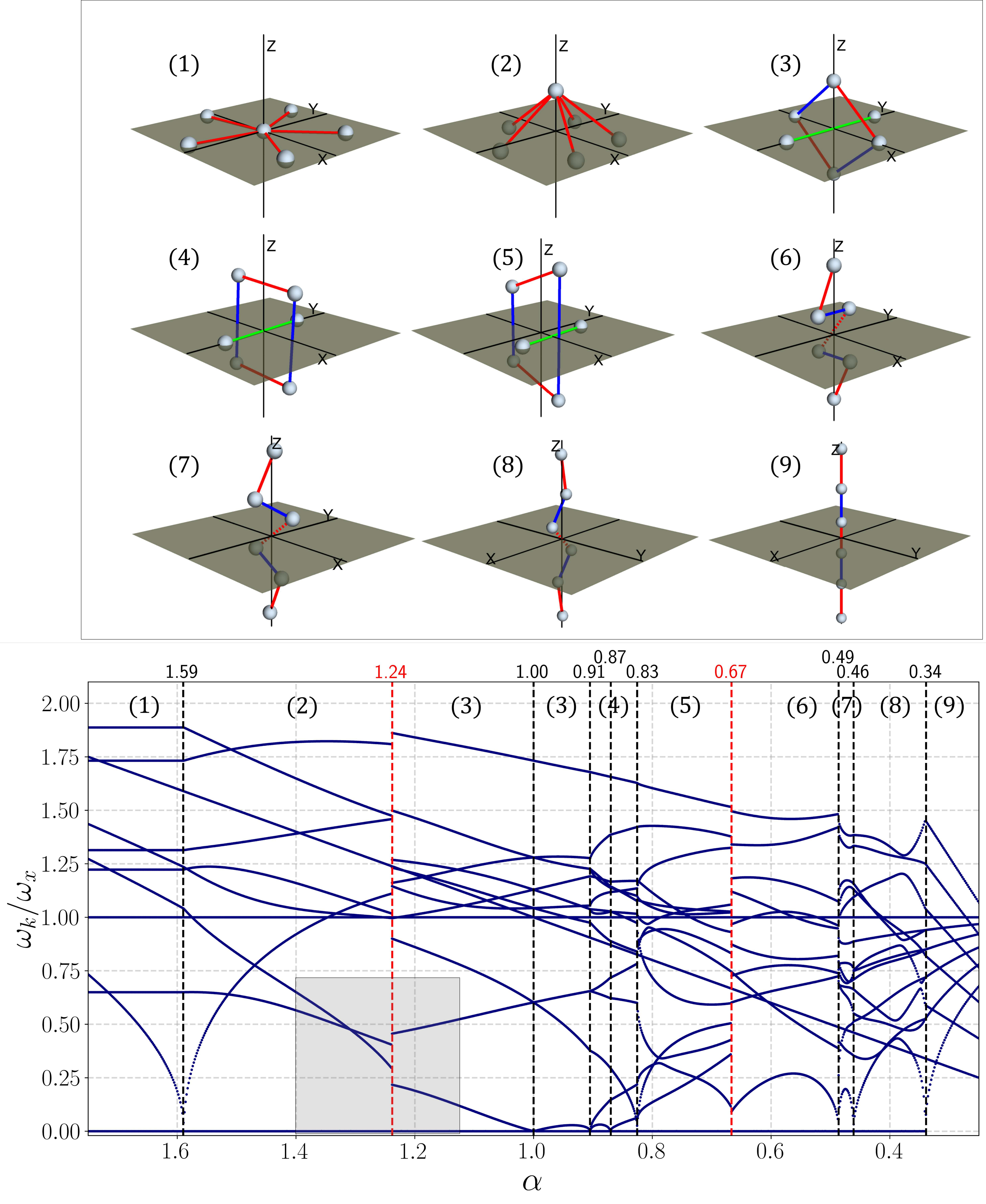}
    \caption{\label{fig:6ion-cmenm} \textbf{(Top)} All the 6-ion minimum energy configurations. \textbf{(Bottom)} Corresponding normal-mode frequencies computed at the theoretical limit $T=0$. The grey region is the ROI for the study of stochastic switching between stable configurations.}
\end{figure*}

\end{document}